\documentclass[apjl]{emulateapj}

\usepackage{graphicx}
\usepackage{t1enc}
\usepackage[varg]{txfonts}
\usepackage{epsfig}
\usepackage{rotating}
\usepackage{natbib}
\usepackage{color}

\newcommand{\lbv} {G79.29+0.46}
\newcommand{\irdc} {G79.3+0.3}

\newcommand{\Trot}  {T_\mathrm{rot}}
\newcommand{\mum}   {$\mu$m}
\newcommand{\kms}   {km~s$^{-1}$}

\newcommand{\cmt}   {cm$^{-3}$}
\newcommand{\jpb}   {$\rm Jy~beam^{-1}$}    

\newcommand{\mo}    {$M_{\sun}$}

\newcommand{\nh}    {NH$_3$}

\newcommand{\ctht}   {$c$-C$_3$H$_2$}

\newcommand{\et}    {et al.}
\newcommand{\eg}    {e.\,g.,}

\definecolor{RED}{rgb}{1.0,0.0,0.0}

\shorttitle{}
\shortauthors{Palau et al.}

\begin{document}

\title{A Luminous Blue Variable Star Interacting with a Nearby Infrared Dark Cloud}


\author{Aina Palau\altaffilmark{1},
J. Ricardo Rizzo\altaffilmark{2},
Josep M. Girart\altaffilmark{1},
Christian Henkel\altaffilmark{3, 4}
}
\altaffiltext{1}{Institut de Ci\`encies de l'Espai (CSIC-IEEC), Campus UAB -- Facultat de Ci\`encies, Torre C5 -- parell 2, E-08193 Bellaterra, Catalunya, Spain}
\email{palau@ieec.uab.es}
\altaffiltext{2}{Centro de Astrobiolog\'{\i}a (INTA-CSIC), Ctra. M-108, km~4, E-28850 Torrej\'on de Ardoz, Madrid, Spain}
\altaffiltext{3}{Max-Planck-Institut f\"ur Radioastronomie, Auf dem H\"ugel 69, D-53121 Bonn, Germany }
\altaffiltext{4}{Astronomy Department, King Abdulaziz University, P.O. Box 80203, Jeddah 21589, Saudi Arabia}

\begin{abstract}
G79.29+0.46 is a nebula created by a Luminous Blue Variable (LBV) star candidate characterized by two almost circular concentric shells. 
In order to investigate whether the shells are interacting with the infrared dark cloud (IRDC) G79.3+0.3 located at the southwestern border of the inner shell, we conducted Jansky Very Large Array observations of \nh(1,1), (2,2) and \ctht, and combined them with previous Effelsberg data. The overall \nh\ emission consists of one main clump, named G79A, elongated following the shape of the IRDC, plus two fainter and smaller cores to the north, which spatially match the inner infrared shell. We analysed the \nh\ spectra at each position with detected emission and inferred linewidth, rotational temperature, column density and abundance maps, and find that:
i) the linewidth of \nh(1,1) in the northern cores is 0.5~\kms, slightly larger than in their surroundings; 
ii) the \nh\ abundance is enhanced by almost one order of magnitude towards the northwestern side of G79A;
iii) there is one `hot slab' at the interface between the inner infrared shell and the \nh\ peak of G79A; 
iv) the western and southern edges of G79A present chemical differentiation, with \ctht\ tracing more external layers than \nh, similar to what is found in PDRs.
Overall, the kinematics and physical conditions of G79A are consistent with both shock-induced and UV radiation-induced chemistry driven by the LBV star.  
Therefore, the IRDC is not likely associated with the star-forming region DR15, but located farther away, near G79.29+0.46 at 1.4~kpc.
\end{abstract}


\section{Introduction \label{sint}}

Luminous Blue Variable (LBV) stars are massive stars undergoing ejection events before their explosion as a Type\,II supernova (\eg\ Groth et al. 2013; Vamvatira-Nakou \et\ 2013). Although understanding massive star evolution is crucial in our understanding of galactic-scale phenomena, the details of the mass-loss episodes during the LBV phase are not well known yet. This is because to properly model these ejection events, it is required a basic knowledge of the properties of the interstellar medium surrounding the LBV star. However, only a few LBV stars have been studied in detail together with their surrounding medium (\eg\ Rizzo \et\ 2001; Loinard \et\ 2012). 


\begin{figure}
\begin{center}
\begin{tabular}[b]{c}
         \epsfig{file=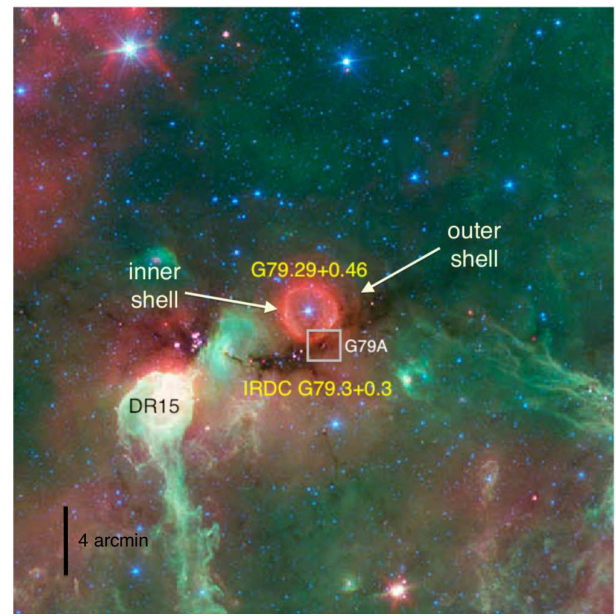,  width=8.cm, angle=0} \\
\end{tabular}
\caption{
General view of the G79 region from 4-color composite Spitzer/IRAC+MIPS image (blue: 3.6~\mum, blue-green: 4.5~\mum, green: 8.0~\mum, red: 24~\mum; NASA/JPL-Caltech/Harvard-Smithsonian CfA). The LBV star is located at the center of the field, with its inner shell being the most outstanding feature at 24~\mum\ (red). 
The white square marks the region studied in this work.
}   
\label{fglobal}
\end{center}
\end{figure}

\lbv\ is a nebula projected toward the center of the Cygnus\,OB2 association, $\sim10'$ to the northwest of the massive star-forming region DR\,15, and  $\sim1'$ to the north of the infrared dark cloud \irdc\ (hereafter the IRDC, see Fig.~\ref{fglobal}).
The distance to \lbv\ is adopted to be $\sim1.4$~kpc (Rizzo \et\ 2014, referred to as `Paper I').
\lbv\ has two almost-circular concentric shells surrounding the star, with the inner one emitting from the mid-infrared (Jim\'enez-Esteban \et\ 2010) up to the centimeter range (Higgs \et\ 1994; Umana \et\ 2011). 
In addition, G79.29+0.46 harbors the LBV candidate with one of the clearest and best studied shells of CO gas (e. g., Rizzo et al. 2008; Petriella et al. 2012; Paron et al. 2012), and is the only candidate with a molecular shell detected in \nh\ (Paper I). Thus, \lbv\ seems to have formed in a particularly dense cloud which was not fully disrupted soon after the birth of the LBV star. 
What is more, the southwestern side of the molecular shell shows hints of interaction with the surrounding medium. Rizzo et al. (2008) report shocked CO in the southwestern side of the shell, at the position of the clump named G79A, where Umana \et\ (2011) also suggest interaction between the shell and the surrounding dense gas, based on the morphology of the centimeter emission at that position. 
Very recently, in Paper I, we report an \nh\ over-abundance towards the northern edge of G79A, and suggest that it could be produced by 
the passage of a low-velocity shock. Thus, G79A turns to be one of the best laboratories to study the direct interaction of LBV ejecta and the surrounding interstellar medium. In this letter, we combine the Effelsberg observations presented in Paper I with new interferometric data in G79A, and provide a close-up view of the physical processes involved in such an interaction.

\section{Observations\label{sobs}}

We used the Jansky VLA\footnote{The Very Large Array (VLA) is operated by the National Radio Astronomy Observatory (NRAO), a facility of the National Science Foundation operated under cooperative agreement by Associated Universities, Inc.} (JVLA) on 21 July 2010, and 9 and 14 Aug 2010 to observe the metastable lines $(J,K)=$\,(1,1) and (2,2) of \nh\ (first and third days) and the $J_\mathrm{K,Ka}=$1$_{1,0}$--1$_{0,1}$
 rotational transition of \ctht, under the project AP581, with the array in the D configuration. The projected baselines ranged from 40 to 1010~m. The phase center of the observations was (RA, Dec)$_\mathrm{J2000}$=(20:31:37.90, +40:19:59.0), and the FWHM of the primary beam at the frequency of the observations is $2.2'$.
\nh(1,1) and (2,2) were observed simultaneously in full polarization mode, with spectral windows of 1.0~MHz and a channel spacing of 15.625~kHz (0.20~\kms; rest frequencies 23694.495 and 23722.633~MHz). This allowed us to cover the main line and one inner satellite of the (1,1) transition, and the main line of the (2,2) transition. \ctht\ was observed in dual polarization, with a spectral window of 2.0~MHz and a channel spacing of 7.813~kHz (0.13~\kms; rest frequency 18343.143~MHz).
Data were calibrated using the software CASA (McMullin \et\ 2007), following the standard calibration procedures. 
Passband response was obtained from observations of 3C84, and phase calibration was done by observing the quasar J2015+3710, for which we measured bootstrapped fluxes of $3.26\pm0.03$~Jy, $3.29\pm0.01$~Jy, and $3.82\pm0.03$~Jy on the three observing days. The third day suffered from a flux calibration problem yielding an uncertainty in the absolute flux scale of 20\%. 
Imaging and combination with Effelsberg data (from Paper I) was done using MIRIAD (Sault et al. 1995) for a taper of $10''$ and sampling the zero-spacing $uv$-hole with 500 visibility points generated with the single-dish data. The final rms of the combined \nh(1,1) image is 9~m\jpb, and the synthesized beam is $14\farcs63\times10\farcs57$, with P.A.=51\fdg16. For the \nh(2,2) image the rms is 7~m\jpb, and the beam is $15\farcs45\times10\farcs47$, P.A.=51\fdg56.
For \ctht\ 
the final JVLA-only cleaned images have a synthesized beam of $9\farcs04\times8\farcs05$, P.A.=19\fdg76, and an rms per channel of 3~m\jpb.

%
%

\begin{figure}
\begin{center}
\begin{tabular}[b]{c}
    \epsfig{file=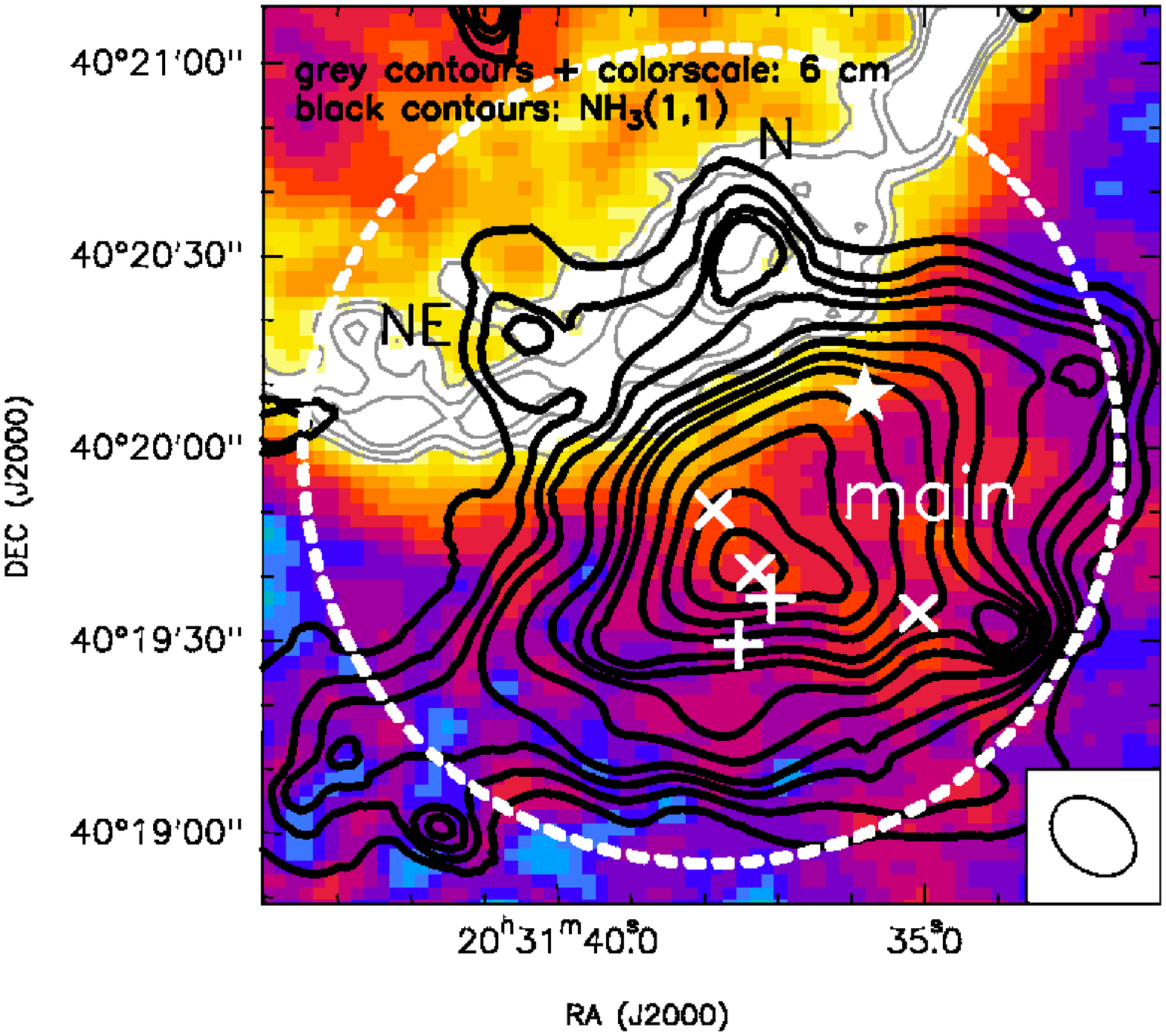,  height=8.5cm, angle=270} \\
    \epsfig{file=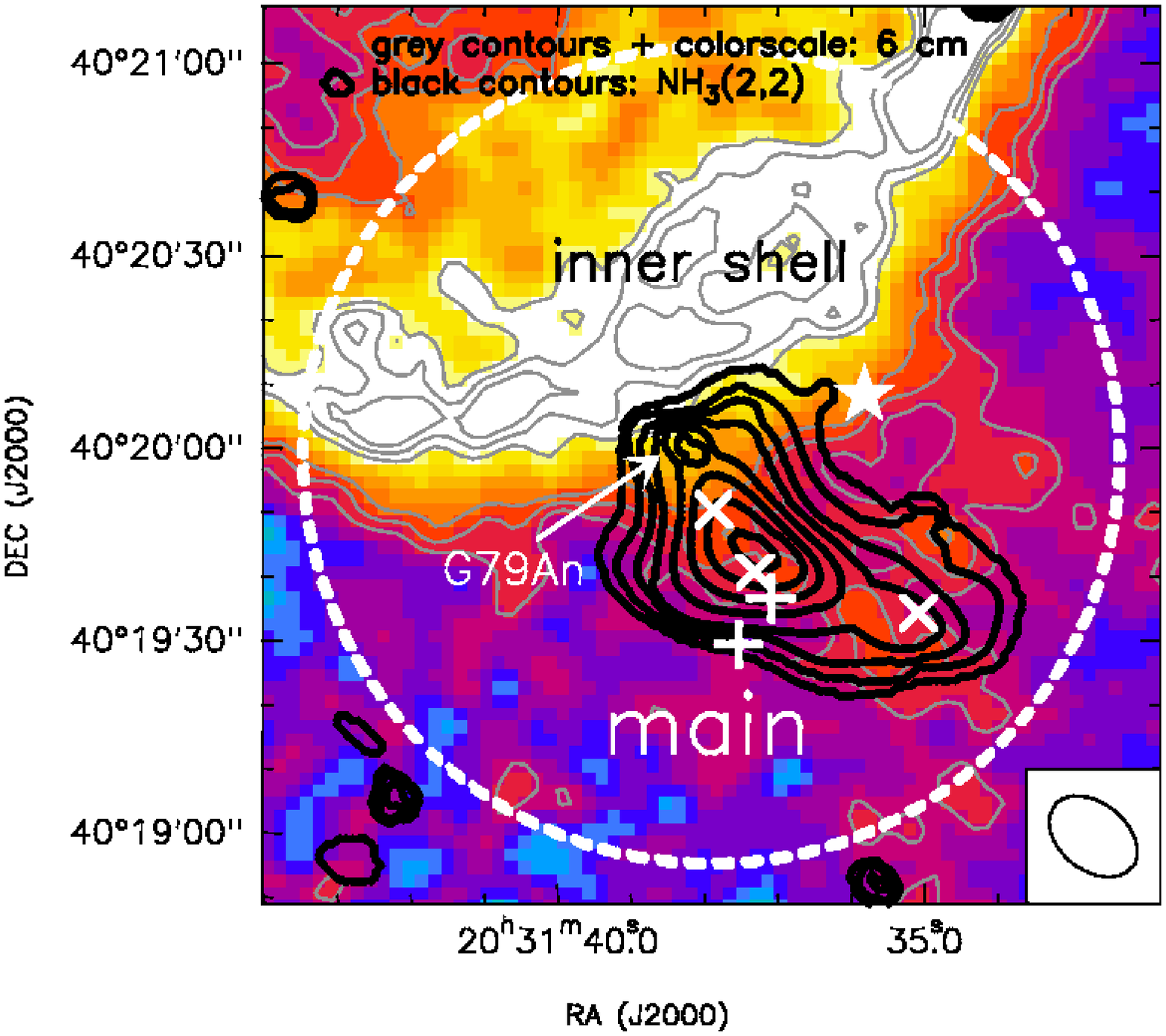,  height=8.5cm, angle=270} \\
\end{tabular}
\caption{
{\it Top:} 
Black contours: zero-order moment of the combined JVLA+Effelsberg \nh(1,1) emission integrated over all the velocity range. 
{\it Bottom:} The same for the (2,2) line.  
In both panels, the white circle indicates the primary beam of the JVLA observations, and the colorscale and grey contours correspond to the 6~cm emission from Higgs \et\ (1994). 
Black contours are 5, 10, 15, 20, 30, 40, 50, 60, 80, 100, 120, and 140 times 8 (1.6)~\jpb\,\kms\ for \nh(1,1) (\nh(2,2)).
Grey contours are (1.25, 1.35, 1.45) 2.0, 2.2, 2.4~K for (bottom) top panel.
The star symbol indicates the position of the shock reported by Rizzo \et\ (2008), the plus signs correspond to 8~\mum\ Spitzer sources, and the tilted crosses correspond to 70~\mum\ Herschel sources. The central and westernmost tilted crosses coincide with 1.2~mm continuum peaks (Motte et al. 2007).
}   
\label{fnh3m0}
\end{center}
\end{figure}

\begin{figure*}
\begin{center}
\begin{tabular}[b]{cc}
         \epsfig{file=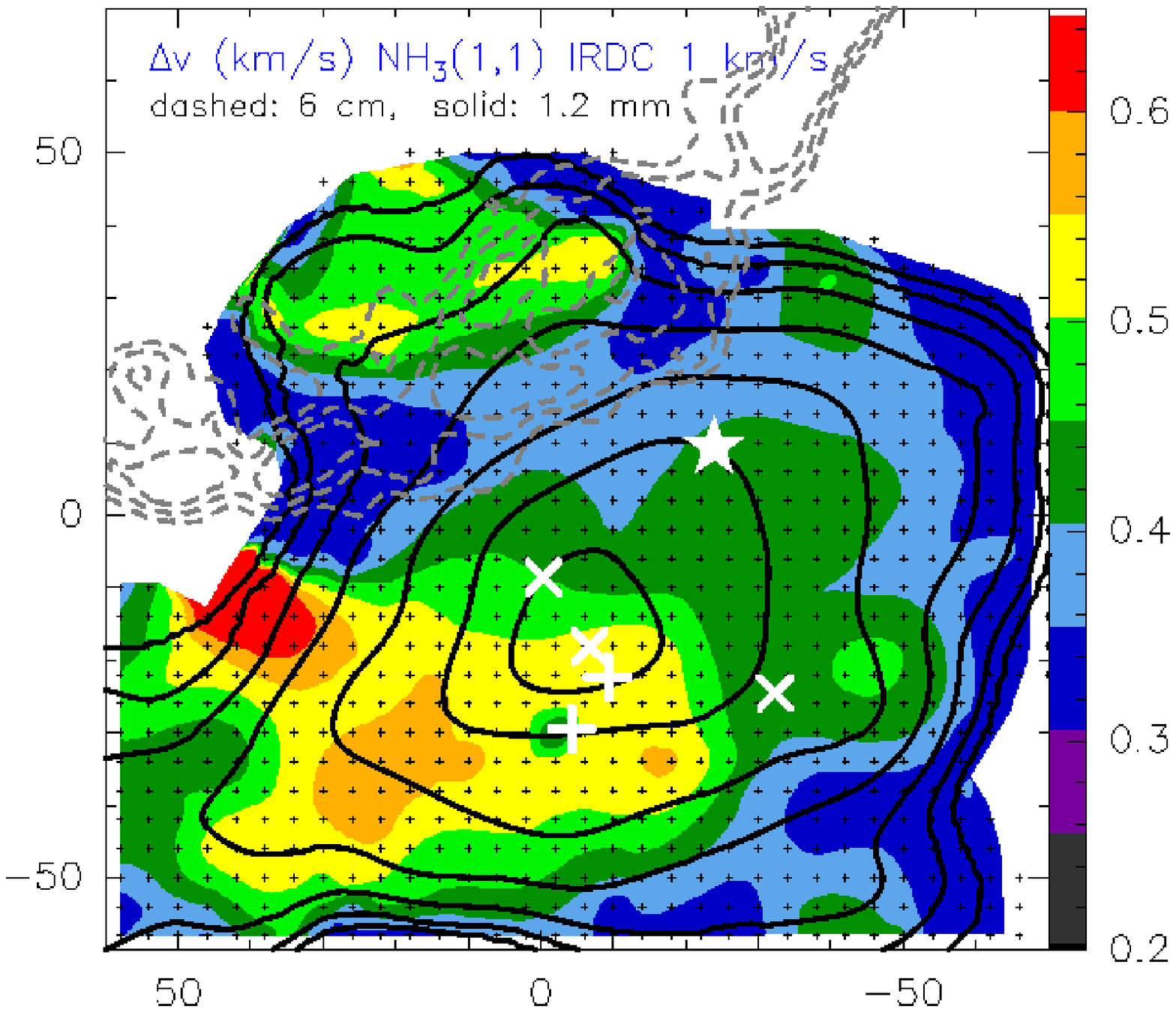,  width=7.0cm, angle=270} &
         \epsfig{file=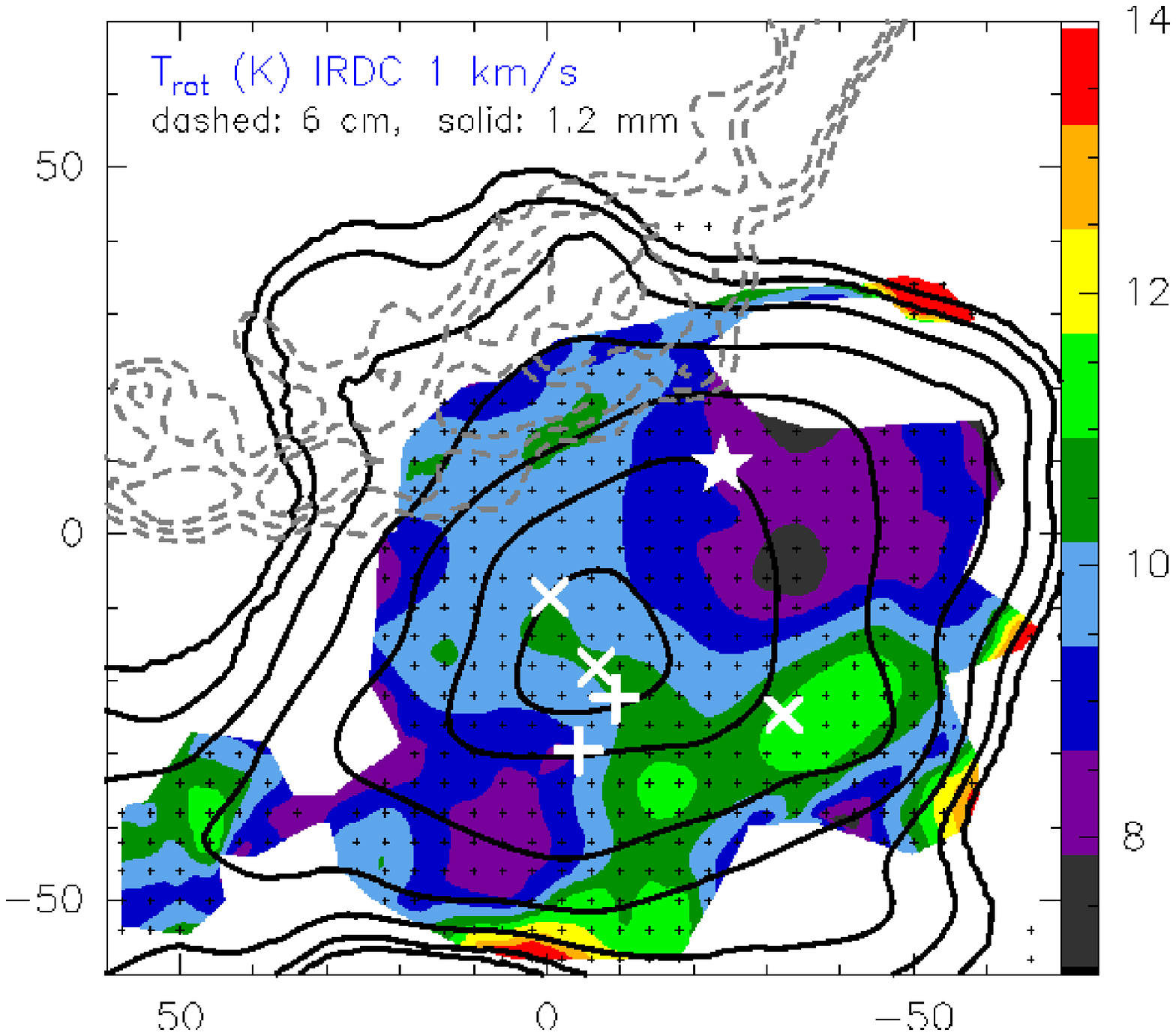,  width=7.0cm, angle=270} \\
         \epsfig{file=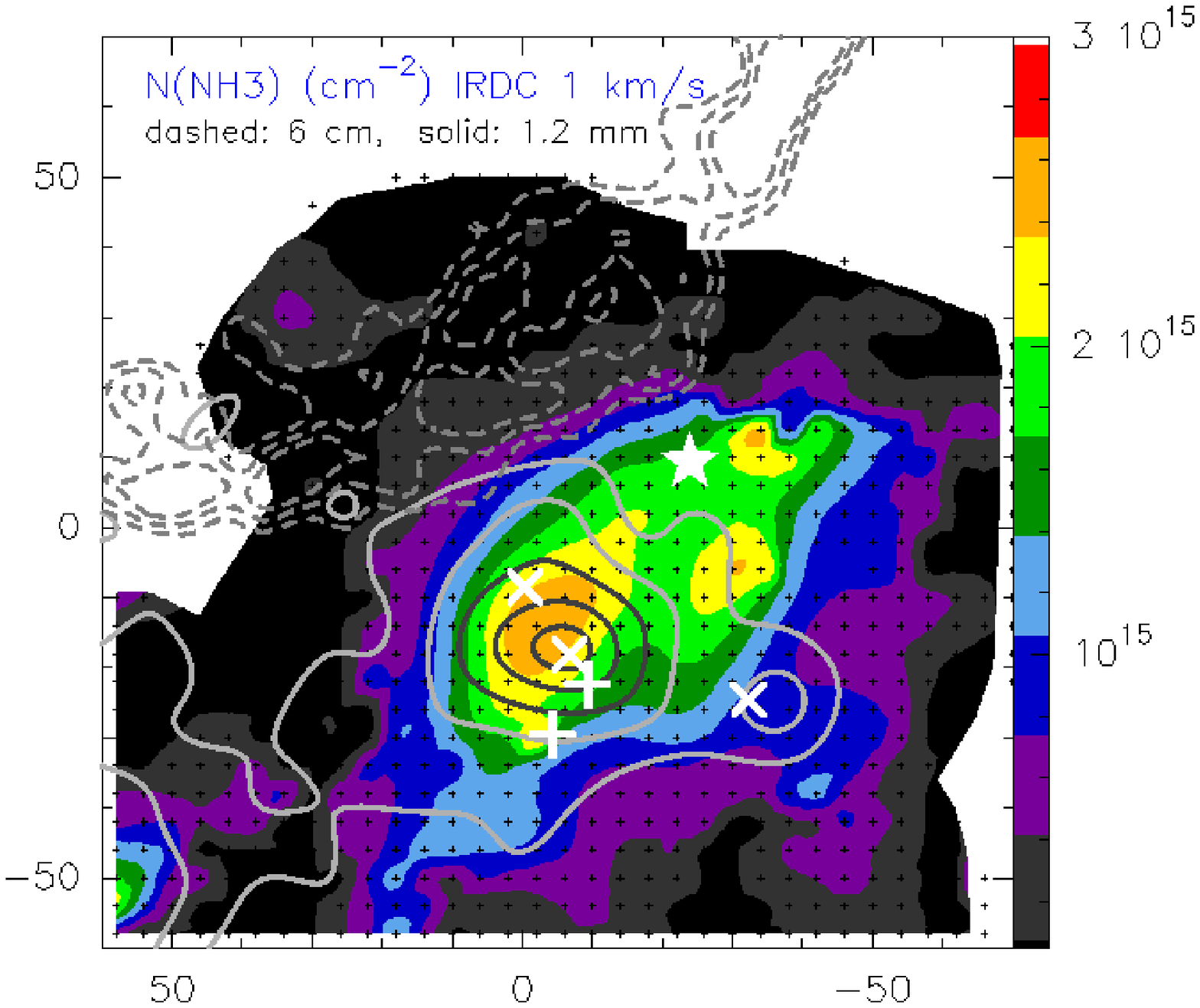,  height=8.5cm, angle=270} &
         \epsfig{file=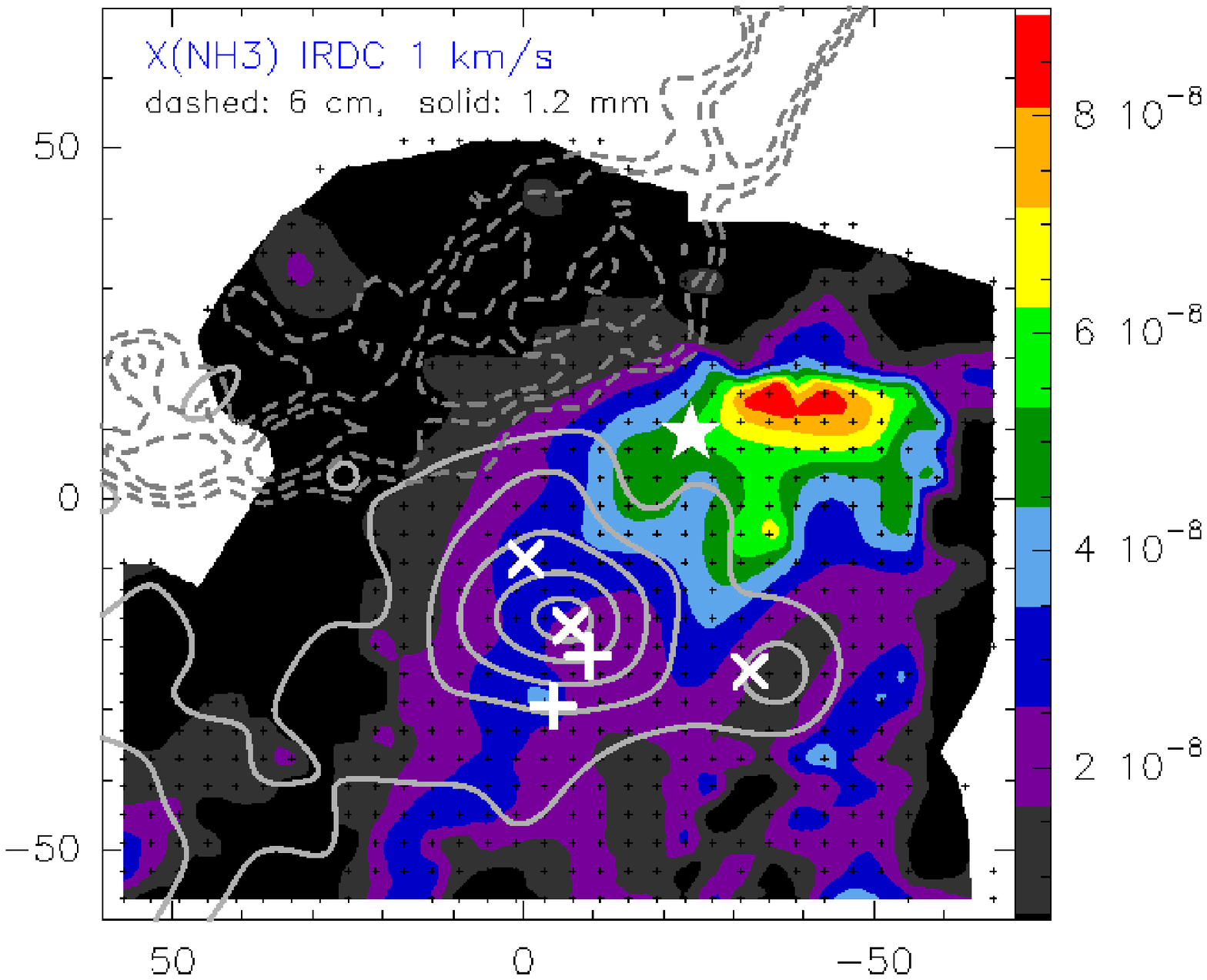,  height=8.5cm, angle=270} \\
\end{tabular}
\caption{Inferred \nh\ parameters for the combined maps Effelsberg + JVLA convolved to $15''$, for the velocity component at 1~\kms\ (corresponding to the IRDC). 
{\it Top-left:} Linewidth distribution of \nh(1,1).
{\it Top-right:} $\Trot$.
{\it Bottom-left:} \nh\ column density. 
{\it Bottom-right:} \nh\ abundance. 
In the two top panels, solid contours correspond to the zero-order moment of the \nh(1,1) channel map convolved to $15''$ (contours are 4, 8, 16, 32, 64, 128, and 200 times 5~\jpb\,\kms). In the bottom panels, solid contours correspond to 1.2~mm continuum emission from Motte et al. (2007) convolved to $15''$ (contours are 9, 12, 15, 18, and 20 times 8.7~m\jpb). In all panels, dashed contours correspond to 6~cm continuum emission as in Fig.~2. Small plus signs indicate the positions where spectra have been fitted and analysed (Sect.~3). Symbols as in Fig.~\ref{fnh3m0}. Coordinates are offsets in arcseconds with respect to the phase center of the JVLA observations (Sect.~2), which is located at ($-50\farcs08, -119\farcs81$) with respect to the LBV star.
}   
\label{fphyspar}
\end{center}
\end{figure*}

\section{Methodology \label{smeth}}

In order to infer the physical parameters from the \nh\ (Effelsberg+JVLA) emission we extracted \nh(1,1) and (2,2) spectra in a regular grid of $4''\times4''$ covering all the \nh(1,1) emission and fitted the hyperfine structure of the \nh(1,1) emission for those spectra with S/N larger than 4,  and fitted a gaussian profile to the \nh(2,2) line for those spectra with S/N larger than 3. All the spectra present emission at $v_\mathrm{LSR}\sim1$~\kms\ (approximately ranging from 0.5 to 1.5~\kms), but some particular spectra show additional velocity components in the range $-2.5$ to 2.5~\kms. By following the same methodology described in detail in the Appendix of Busquet \et\ (2009), we inferred the velocity, linewidth, opacity of the main line (ranging from $\la1$ to 3), total column density (corrected for the primary beam response) and rotational temperature $\Trot$ for each point of the grid, for the velocity component at 1~\kms, and built maps for these parameters.
Multiple velocity components in the spectra were fitted individually.
The rotational temperature $\Trot$ was estimated at those positions where both \nh(1,1) and \nh(2,2) were detected. Wherever \nh(2,2) was not detected but \nh(1,1) was, we calculated the upper limit of $\Trot$ to be $\sim12$~K. Since this upper limit is very close to the lowest $\Trot$ typically seen in starless cores ($\sim8$--10~K), by adopting this upper limit for $\Trot$ we provide reasonable \nh\ column density and abundance estimates at the positions with only \nh(1,1) detected. Typical uncertainties in the rotational temperature are around 1--2~K. Note also that $\Trot$ inferred from \nh(1,1) and (2,2) lines is a good approximation to the kinetic temperature for $\Trot\la25$~K  
(\eg\ Walmsley \& Ungerechts 1983; Danby \et\ 1988).

Last, we built an \nh\ abundance map by using our inferred \nh\ column density and estimating the H$_2$ column density from the Motte et al. (2007) 1.2~mm continuum image (after convolving the \nh(1,1), (2,2) channel maps, and the 1.2~mm image to the same beam of $15''$). To convert to H$_2$ column density we assumed a dust temperature of 10~K, an opacity at 1.2~mm of 0.010~cm$^2$\,g$^{-1}$ (Ossenkopf \& Henning 1994), and a gas-to-dust ratio of 100 (see Appendix of Paper I for further details).

\section{Results \label{sres}}

The bulk of the \nh(1,1) emission is found at  $v_\mathrm{LSR}\sim1$~\kms, similar to the systemic velocity of the IRDC (Redman \et\ 2003; Higuchi \et\ 2013). However, an important number of positions, both in \nh(1,1) and (2,2), present additional velocity components ranging from  $-2.5$ to $+2.5$~\kms, consistent with the range where we discovered the \nh\ counterpart of the LBV shell.
%
Fig.~\ref{fnh3m0} presents the distribution of the zero-order moment of the combined (Effelsberg+JVLA) \nh(1,1) and (2,2) data towards G79A. 
The \nh(1,1) emission consists of one main centrally-peaked clump with some substructure being elongated in the (south)east-(north)west direction. The \nh(1,1) clump follows the large-scale elongation of the IRDC at this position (Fig.~1), its peak falls about $\sim30''$ to the south of the inner shell, and is coincident with a 70~\mum\ source and with the peak of 1.2~mm continuum emission (hereafter, the G79A peak; Fig.~\ref{fnh3m0}). 
There are two fainter cores to the north of the main \nh(1,1) clump (labeled NE and N in Fig.~2) which are spatially coincident with the LBV inner shell traced by the 6~cm emission. 
While core N mainly emits at 1~\kms, the NE core presents an additional velocity component at $\sim-0.1$~\kms.
%
Thus, \nh\ emission at the same velocity as the IRDC is clearly found associated with the LBV shell.

The emission of the (2,2) line is less extended than the (1,1). In addition, it is elongated in the northeast-southwest direction, following the faint 6~cm emission seen downstream of the shell. The strongest \nh(2,2) peak falls at the same position as the G79A peak. Interestingly, the \nh(2,2) integrated emission presents a secondary peak about $\sim10''$ to the north of the G79A peak, exactly at the border of the LBV shell (and labeled as G79An; see below for further details). 

In Fig.~\ref{fphyspar} we present the maps of linewidth, $\Trot$, column density and \nh\ abundance for the velocity component at 1~\kms\ (corresponding to the IRDC). 
The linewidth is around 0.3--0.4~\kms\ in almost all the clump, except in two separated regions where it reaches values $\ga0.5$~\kms. One of the regions is mainly associated with the southeastern part of the G79A clump. The other coincides with the NE and N cores, at the shell location.
%
The rotational temperature map reveals a region of enhancement towards the southwest of G79A, with $\Trot$ up to 13~K, which is almost devoid of mid/far-infrared sources, while $\Trot$ decreases down to 8~K towards the northwest and southeast of G79A. The temperature is $\sim10$~K around the G79A peak, where several infrared sources are located. 
%
The \nh\ column density reaches its highest value at the G79A peak, and extends along the southeast-northwest direction.
As for the \nh\ abundances, we find values ranging from $1\times10^{-8}$ to $9\times10^{-8}$,
and interestingly the \nh\ abundance peak is shifted with respect to the G79A peak about $\sim40''$ towards the northwest, close to the position of the shock reported by Rizzo et al. (2008; star symbol in Figs.~2 and 3).

\begin{figure}
\begin{center}
\begin{tabular}[b]{c}
         \epsfig{file=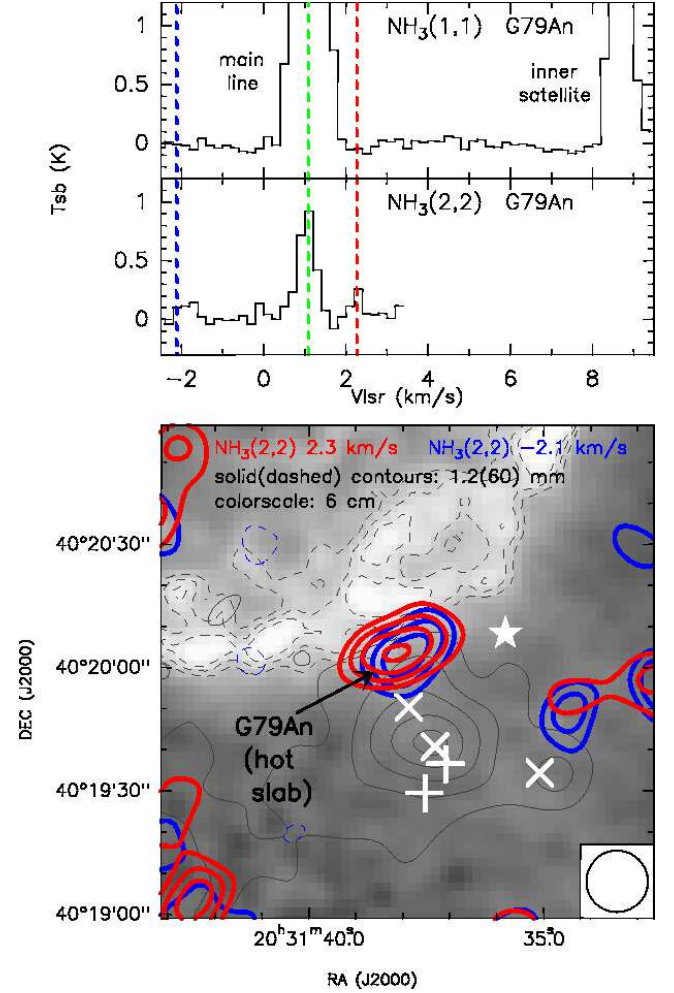,  width=8.5cm, angle=0} \\
\end{tabular}
\caption{
{\it Top:} \nh(1,1) and (2,2) spectra toward G79An (`hot slab'). Rms noise is 0.04 and 0.07~K, respectively.
{\it Bottom:} Channel map at $-2.1$ (blue contours) and 2.3 (red contours)~\kms\ of the \nh(2,2) cube convolved to $15''$ (Effelsberg+JVLA). Contours are $-2$, 2, 3, 4 and 5 times $\sim7$~m\jpb. Dashed contours (6~cm), solid contours (1.2~mm), and symbols as in Fig.~\ref{fphyspar}.
}   
\label{fhotspots}
\end{center}
\end{figure}

As mentioned above, the \nh(1,1) and (2,2) spectra show
multiple velocity components ranging from $-2.5$ to $+2.5$~\kms.
The velocity components (different to 1~\kms\ of the IRDC) with higher intensity are found at 2.3 and $\sim-2$~\kms, and are detected only in the \nh(2,2) line. These velocity components are responsible for the secondary peak G79An (Fig.~\ref{fnh3m0}-bottom). Fig.~\ref{fhotspots} shows the \nh(2,2) channels at $-2.1$ and 2.3~\kms, overplotted on the 6~cm and 1.2~mm emission, and reveals that these velocity components arise from material located in between the G79A peak and the LBV shell. Since G79An is not detected in \nh(1,1), we infer a lower limit of $\Trot$ of $40$~K, much warmer than the typical values measured for the 1~\kms\ component of the IRDC (Fig.~3). In addition, a 2D gaussian fit to G79An at 2.3~\kms\ yields a size of $26''\times14''$ with P.A.=$-68$\degr, which matches the elongation of the shell at this position. These velocity components point to the presence of a `hot slab' at the interface between the LBV shell and G79A, which is not associated with 8 or 70~\mum\ point-sources (Fig.~4). 

\begin{figure}
\begin{center}
\begin{tabular}[b]{c}
     \epsfig{file=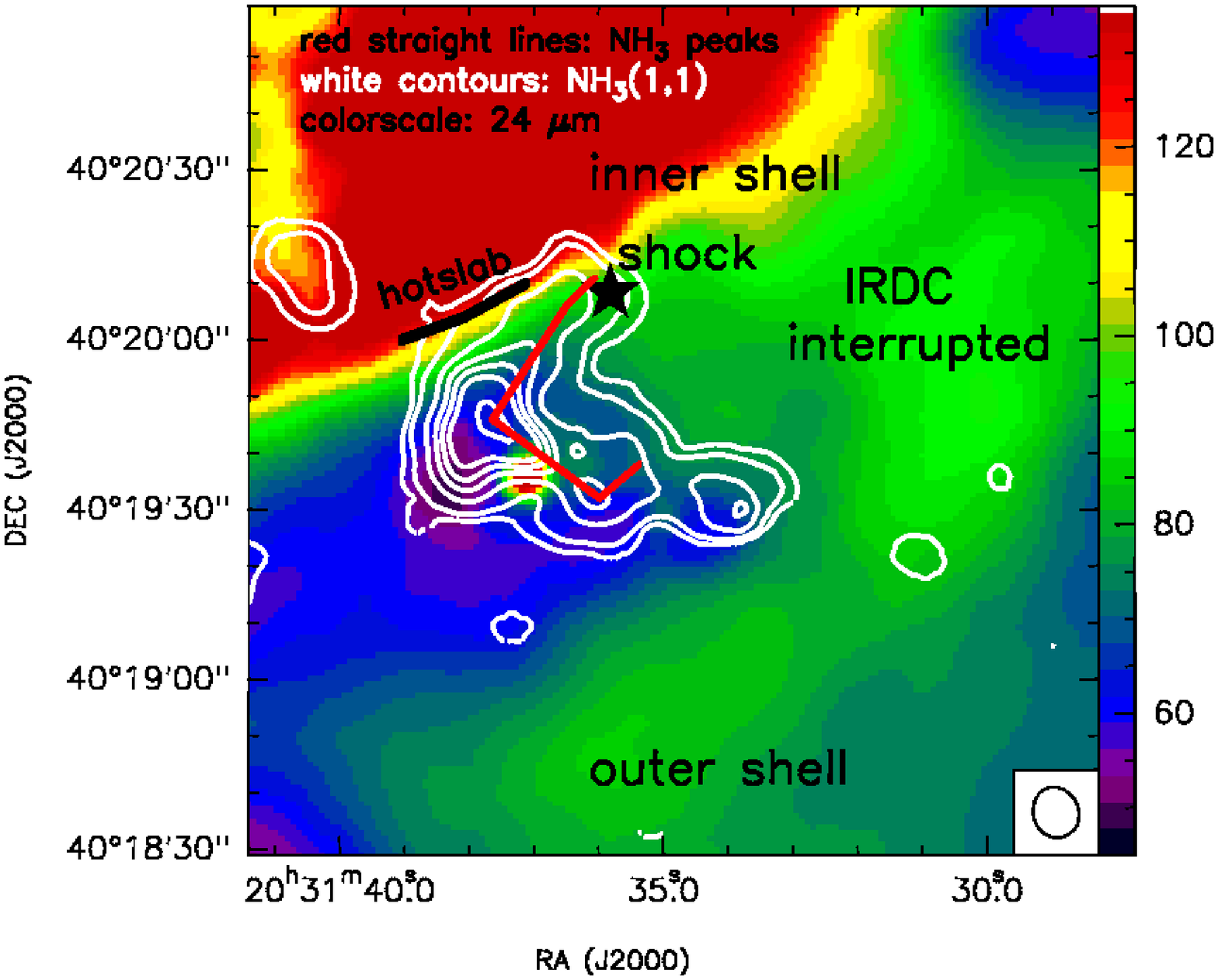,  height=9cm, angle=270} \\
     \epsfig{file=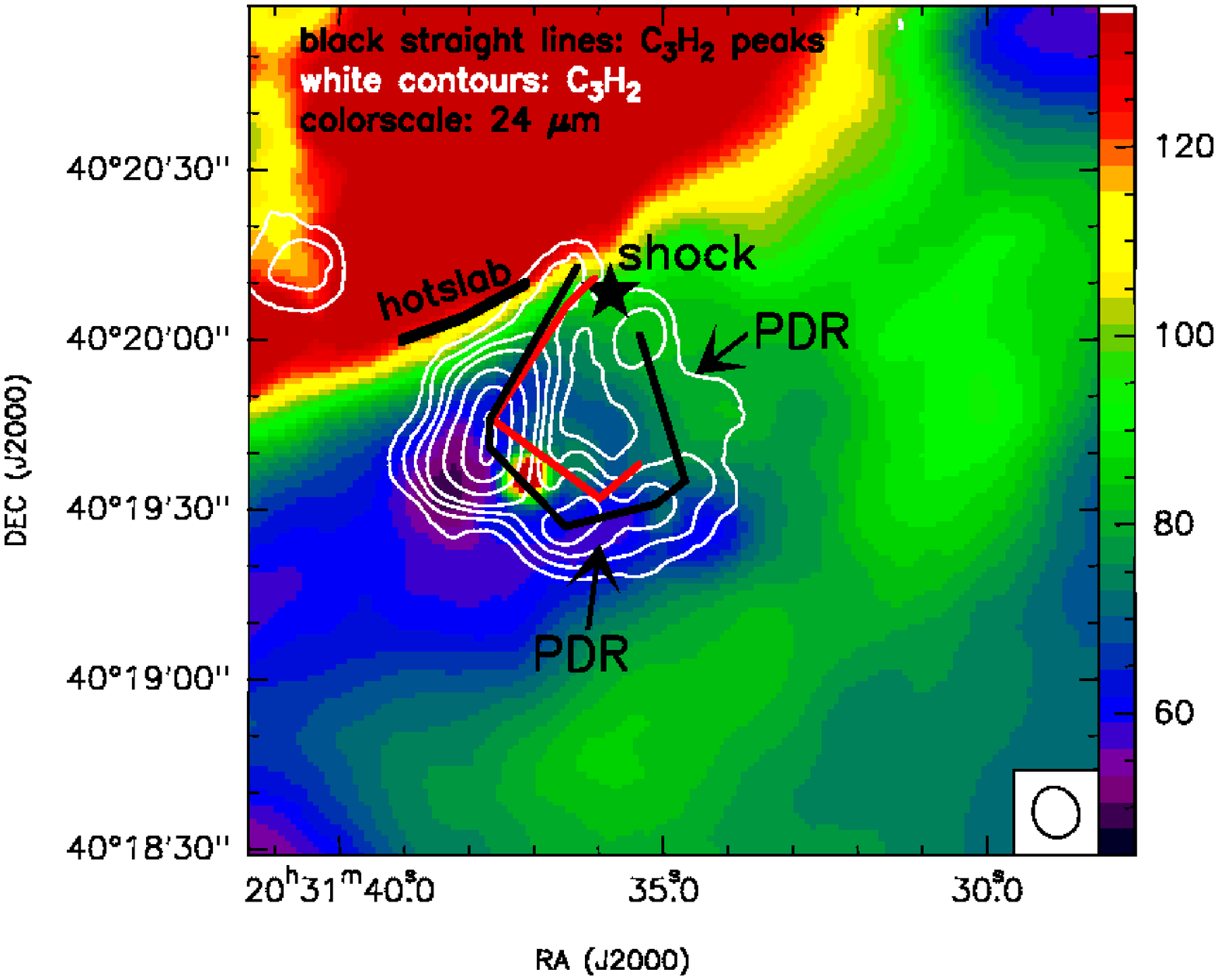,  height=9cm, angle=270} \\
\end{tabular}
\caption{Only JVLA. {\it Top:} Contours: \nh(1,1) zero-order moment. 
{\it Bottom:} Contours: \ctht\ zero-order moment. Black straight lines connect the peaks of the \ctht\ emission. 
%
In both panels, the black star marks the position of the shock reported in Rizzo \et\ (2008); the black thick line marks the `hot slab' (G79An) reported in this work (Fig.~4); the red straight lines connect the peaks of the \nh(1,1) emission, and the color scale is the 24~\mum\ Spitzer image (Jim\'enez-Esteban \et\ 2010). 
}   
\label{fc3h2}
\end{center}
\end{figure}

In Fig.~\ref{fc3h2} we present the emission of \ctht\ obtained using the JVLA data only (bottom panel) overplotted on a Spitzer 24~\mum\ image (Jim\'enez-Esteban et al. 2010). For comparison, we plot the emission of \nh(1,1) using also JVLA data only (top panel). Thus, both datasets are filtering out the same large spatial scales. While the emission of \nh(1,1) is following mainly a `tilted-L' shape, with the northern arm following the inner shell at 24~\mum, the \ctht\ emission is present towards the western side of G79A where no \nh\ is found. 
To compare the \ctht\ emission to the \nh(1,1) emission, in Fig.~5-bottom we connect with black lines the peaks of \ctht, and overplot in red lines the connected peaks of \nh(1,1). The figure shows that the black lines tracing \ctht\ are located farther to the south and to the west as compared to the red lines tracing \nh, and thus \ctht\ seems to trace material from outer layers of the G79A clump.

Finally, we calculated some physical parameters of G79A from the millimeter continuum emission. Using the 1.2~mm data of Motte \et\ (2007), we estimated a mass for G79A of 80---150~\mo, for dust temperatures from 10 to 15 K, and adopting the same opacity and dust-to-gas ratio given in Section~3, and the same distance as the LBV star (1.4~kpc, see Sect.~5). The uncertainty in the masses due to the opacity law is estimated to be a factor of two. For the size, a 2D gaussian fit to the 1.2~mm continuum clump yields a deconvolved size of $44''$ in diameter. This corresponds to a density in G79A of 0.8--$1.5\times10^5$~\cmt.

\section{Discussion and Conclusions \label{sdis}}

The \nh(1,1) and (2,2) and \ctht\ data presented in this work reveal a number of features which definitely confirm an interaction of the infrared shells created by the LBV star with the IRDC. This has been possible because of the combination of both JVLA and Effelsberg data, allowing us to recover information from large ($\sim1$~pc) to small ($\sim0.05$~pc) spatial scales.

%
First, we find an increase of the \nh\ abundance towards the northwest of G79A, of one order of magnitude (up to $9\times10^{-8}$). In Paper I, we report also an abundance increase to the north of G79A, but could not accurately determine its position because we used Effelsberg data alone. In this paper we find that the \nh\ increase takes place precisely at the position where Rizzo \et\ (2008) report a shock in CO\,(3--2). Overall, this suggests that the north(western) part of the G79A clump is being shocked by the expanding shell of the LBV star, releasing \nh\ molecules from the dust grains back to the gas phase. Viti \et\ (2011) model the effects of the propagation of a C-type shock into dense ($\sim10^5$~\cmt) interstellar material, including the sputtering of dust grains, and predict an increase in \nh\ abundance from 10$^{-9}$ up to $\sim10^{-5}$, depending on the particular conditions of the shock and pre-shock material. This \nh\ abundance increase is consistent with our observations, and the physical conditions adopted in the model are also consistent with those in G79A, as the density in G79A is estimated
%
%
to be $\sim10^5$~\cmt\ (Section~4), and the velocity required for efficient sputtering of dust grains is $\ga10$~\kms\ (van Loo \et\ 2013), similar to the shock velocity measured by Rizzo \et\ (2008). 
An increase of \nh\ abundance in cores shocked by the passage of an outflow {\bf is} also observed in star-forming regions, such as HH\,2 (Torrelles \et\ 1992; Girart \et\ 2005).

Further evidence of the gas in G79A being shocked comes from the `hot slab' found near the G79A peak (Fig.~\ref{fhotspots}). The different velocity components and $\Trot$ of this `hot slab' are fully consistent with the Effelsberg spectrum of \nh(3,3) near the same position (Paper I). Thus, the high angular resolution data reported in this work allowed us to locate the origin of the \nh(3,3) hot velocity components discovered with single-dish. The facts that no mid/far-infrared point-sources are associated with the `hot slab', that it is located at the interface between the LBV shell and the G79A peak, that it is elongated along the shell direction, and that their velocities are slightly different to that of the IRDC, all indicate that this `hot slab' is probably produced by shocks generated by the expanding shell from the LBV star.

Furthermore, our study suggests that the interaction of the LBV shell with the G79A clump is not only through shocks, but also radiative. 
This is suggested by a temperature enhancement (up to 12--13~K) extending $\sim40''$ in the south-western part of the G79A clump, which is almost devoid of mid/far-infrared sources and thus could be produced by external heating.
%
Moreover, in this southwestern region of G79A, \ctht\ seems to trace the outer layers of the clump as compared to \nh\ (Fig.~5). As expected in photon-dominated regions (PDRs), \nh\ is photo-destroyed in the outer parts of the PDR, while \ctht\ is known to survive in hot regions ($\sim100$~K) for several 10$^5$ yr (Hassel et al. 2011; Aikawa et al. 2012), as shown also by recent observations of the HorseHead nebula and Mon R2 PDRs (Rizzo \et\ 2005; Gerin \et\ 2009; Pilleri \et\ 2013). The strong 24~\mum\ emission surrounding not only the northern side of G79A but also the western and southern edges (Fig.~5) is consistent with this scenario. 
Interestingly, the outer infrared shell lies exactly to the south of G79A (Jim\'enez-Esteban \et\ 2010; Umana \et\ 2011), and the IRDC is interrupted exactly to the west of G79A (Fig.~5; Paper I), where the circular geometry of the inner shell is distorted  (Umana \et\ 2011). 
Thus, the southern and western edges of the G79A clump seem to be bathed by (mid-infrared) radiation probably coming also from the LBV star.
\\


The characterization of the dense gas surrounding the G79.29+0.46 nebula, as well as the presented details of the interaction of the nebula with the IRDC, constitute an excellent base for future modeling of the mass-loss episodes of the underlying LBV star. 
In addition, in Paper I we show that the IRDC is probably located in between the DR15 massive-star forming region and the LBV nebula G79.29+0.46. The results presented in this work show for the first time clear evidence that the LBV ejecta and the IRDC are interacting. Thus, the IRDC G79.3+0.3 is not likely associated with the DR15 star-forming region at 800~pc, as usually assumed in the literature, but close to G79.29+0.46 at $\sim1.4$~kpc.




\acknowledgments
\begin{small}
A.P. is grateful to Gemma Busquet for useful hints about combination of Effelsberg + JVLA data.  A.P. and J.M.G. are supported by the Spanish MICINN grant AYA2011-30228-C03-02 (co-funded with FEDER funds), and by the AGAUR grant 2009SGR1172 (Catalonia). J.R.R. acknowledges support from MICINN (Spain) grants CSD2009-00038, AYA2009-07304, and AYA2012-32032. 
\end{small}



\end{document}